\documentclass[12pt]{article}
\usepackage{latexsym}
\usepackage{amssymb}
\usepackage[english]{babel}
\usepackage{amsmath}
\title{Universal factorized formula for the cross-section of two-particle scattering}
\author{V.I. Kuksa}
\date{Institute of Physics, Southern Federal University,\\
 pr. Stachki 194, Rostov-on-Don, 344090 Russia,\\
 E-mail address: kuksa@list.ru}
\begin{document}

\maketitle
\begin{abstract}
 We analyze the process of two-particle scattering with unstable particle in
an intermediate state. It was shown that the cross-section can be
represented in the universal factorized form for an arbitrary set
of particles. Phenomenological analysis of factorization effect is
fulfilled.

PACS number(s): 11.10St, 130000


\end{abstract}

\pagenumbering{arabic} \setcounter{page}{1}

\section{Introduction}
The peculiar properties of the unstable particles (UP) and
resonances were being discussed during the last decades. Among them,
the assumption that the decay of UP or resonance (R) proceeds
independently of its production remains of interest \cite{1,2,3}.
Formally, this effect is expressed as the factorization of a
cross-section or decay rate \cite{3}. The processes of type
$ab\rightarrow Rx\rightarrow cdx$ were considered in
Ref.~\cite{3}. It was shown, that the factorization always is valid
for a scalar $R$ and does not take place for a vector and spinor $R$.
The factorization usually is related with the narrow-width
approximation (NWA) \cite{4}, which makes five critical
assumptions \cite{5}.

There is another way to get factorization effect, which is connected
with propagator structure \cite{6}. The decay processes of type $a\rightarrow Rx\rightarrow cdx$,
where $R$ is UP with a large width, were analyzed systematically in
Ref.~\cite{6}. It was shown in this work, that the factorization always is valid
for a scalar $R$, while for a vector and spinor $R$ it occurs
when the propagators' numerators are
$\eta_{\mu\nu}(q)=g_{\mu\nu}-q_{\mu}q_{\nu}/q^2$ and
$\hat{\eta}(q)=\hat{q}+q$, respectively, where
$\hat{q}=q_i\gamma^i$ and $q=\sqrt{q_iq^i}$. Such a structure of
propagators always provides the exact factorization for any tree
process and is an analog of NWA, which is discussed in Section 3.
These propagators were constructed in the model of UP with a smeared mass \cite{7} and
describe some effective (dressed by self-energy insertion) unstable fields.
Note that the structure of the expressions $\eta_{\mu\nu}(q)$ and $\hat{\eta}(q)$
is not related with the choice of the gauge (see the third section).

In this work, we systematically analyze the processes of type
$ab\rightarrow R\rightarrow cd$, where $R$ is scalar, vector
or spinor UP with a large width (or resonance) and $a, b, c, d$
are the stable or long-lived particles of any kind. It was shown
that the cross-section $\sigma(ab\rightarrow R\rightarrow cd)$ can be
represented in the universal factorized form when the same expressions
$\eta_{\mu\nu}(q)=g_{\mu\nu}-q_{\mu}q_{\nu}/q^2$ and
$\hat{\eta}(q)=\hat{q}+q$ are used to describe the propagator's
numerator of vector and spinor UP, respectively. This
result have been received strictly by direct calculations for all
types of particles $a, b, c, d$ and $R$ (Section 2). The factorization approach
is applied in Section 3 for the complicate processes of scattering with the consequent decays of
the final states. In Section 4, we analyze some methodological and phenomenological
aspects of factorization.

\section{Universal factorized formula for the\\ cross-section
of two-particle scattering}

In this section, we consider inelastic scattering of type
$ab\rightarrow R\rightarrow cd$, where $R$ is the UP with a
large width in $s$-channel and $a, b, c, d$ are stable
(quasi-stable) particles of any kind. The vertexes are defined by
the Lagrangian in the simplest standard form:
\begin{align}
 L_k=&g\phi\phi_1\phi_2;\,\,\,g\phi\bar{\psi}_1 \psi_2;\,\,\,g\phi
 V_{1\mu}V^{\mu}_2;\,\,\,gV_{\mu}(\phi_1^{,\mu}\phi_2-\phi_2^{,\mu}\phi_1);\,\,\,
 gV_{\mu}\bar{\psi}_1\gamma^{\mu}(c_V+c_A\gamma_5)\psi_2;\notag\\&gV_{1\mu}V_{2\nu}V_{\alpha}
 [g^{\mu\nu}(p_2-p_1)^{\alpha}+g^{\mu\alpha}(2p_1+p_2)^{\nu}-g^{\nu\alpha}(p_1+2p_2)^{\mu}].
 \label{E:2.1}
\end{align}
In the expressions (\ref{E:2.1}) $\phi, V$ and $\psi$ are the
scalar, vector and spinor fields, respectively, $p_1$ and $p_2$
are the momenta of the particles $a$ and $b$ (or $c$ and $d$).

Here we show, that the cross-section $\sigma(ab\rightarrow R\rightarrow
cd)$ can be expressed in a factorized universal form in terms of
decay widths $\Gamma(R\rightarrow ab)$ and $\Gamma(R\rightarrow
cd)$, when the expressions for propagators' numerators
$\eta_{\mu\nu}(q)=g_{\mu\nu}-q_{\mu}q_{\nu}/q^2$ and
$\hat{\eta}(q)=\hat{q}+q$ are used. This expressions are constructed within the model of unstable particles
with a smeared mass, which briefly considered in Appendix. The validity of these
expressions have been discussed in Refs.~\cite{6,7} and will be
considered in the third section. It is convenient
to employ the universal expressions for widths
$\Gamma(R\rightarrow ab)$ and $\Gamma(R\rightarrow cd)$ in a
stable particle approximation \cite{6}:
\begin{equation}
 \Gamma_i(R\rightarrow
 ab)=\frac{g^2}{8\pi}\bar{\lambda}(m_a,m_b;m_R)f_i(m_a,m_b;m_R),
\label{E:2.2}
\end{equation}
where $m_R^2=q^2, q^2=(p_1+p_2)^2$ and:
\begin{equation}
 \bar{\lambda}(m_a,m_b;m_R)=[1-2\frac{m^2_a+m^2_b}{m^2_R}+\frac{(m^2_a-m^2_b)^2}{m^4_R}]^{1/2}.
\label{E:2.3}
\end{equation}
The same expressions and relations are in order for the width
$\Gamma(R\rightarrow cd)$. The functions $f_i(m_a,m_b;m_R)$ are
defined by the corresponding vertexes. If these vertexes are
described by Eqs.(\ref{E:2.1}), then the functions $f_i$ (further
we omit the arguments) in tree approximation are defined by the following
expressions \cite{6}:
\begin{align}
 &\phi\rightarrow
 \phi_1\phi_2,\,\,f_1=\frac{1}{2m_{\phi}};\,\,\,\,\phi\rightarrow V_1V_2,\,\,
 f_2=\frac{1}{m_{\phi}}[1+\frac{(m^2_{\phi}-m^2_1-m^2_2)^2}{8m^2_1m^2_2}];\notag\\
 &\phi\rightarrow
 \bar{\psi}_1\psi_2,\,\,f_3=m_{\phi}[1-\frac{(m_1+m_2)^2}{m^2_{\phi}}];\,\,\,
 \phi\rightarrow\phi_1V,\,\,f_4=\frac{m^3_{\phi}}{2m^2_V}\bar{\lambda}^2(m_1,m_V;m_{\phi});\notag\\
 &V\rightarrow\phi_1\phi_2,\,\,f_5=\frac{m_V}{6}\bar{\lambda}^2(m_1,m_2;m_V);\,\,\,\,
 V\rightarrow V_1\phi,\,\,f_6=\frac{1}{3m_V}[1+\notag\\ &\,\,\,+\frac{(m^2_V+m^2_1-m^2_{\phi})^2}{8m^2_Vm^2_1}];\notag\\
 &V\rightarrow\bar{\psi}_1\psi_2,\,\,f_7=\frac{2}{3}m_V\{c_{+}[1-\frac{m^2_1+m^2_2}{2m^2_V}-
 \frac{(m^2_1+m^2_2)^2}{2m^4_V}]+3c_{-}\frac{m_1m_2}{m^2_V}\};\notag\\
 &V\rightarrow V_1V_2,\,\,f_8=\frac{m^5_V}{24m^2_1m^2_2}[1+8(\mu_1+\mu_2)-2(9\mu^2_1+16\mu_1\mu_2+9\mu^2_2)+
 8(\mu^3_1-\notag\\ &4\mu^2_1\mu_2
 -4\mu_1\mu^2_2+\mu^3_2)+\mu^4_1+8\mu^3_1\mu_2-18\mu^2_1\mu^2_2+
 8\mu_1\mu^3_2+\mu^4_2],\,\,\mu_{1,2}=m^2_{1,2}/m^2_V;\notag\\
 &\psi\rightarrow
 \phi\psi_1,\,\,f_9=\frac{m_{\psi}}{2}(1+2\frac{m_1}{m_{\psi}}+\frac{m^2_1-m^2_{\phi}}{m^2_{\psi}});\notag\\
 &\psi\rightarrow
 V\psi_1,\,\,f_{10}=m_{\psi}\{c_{+}[\frac{(m^2_{\psi}-m^2_1)^2}{2m^2_{\psi}m^2_V}+\frac{m^2_{\psi}+m^2_1
 -2m^2_V}{2m^2_{\psi}}]-3c_{-}\frac{m_1}{m_{\psi}}\};\notag\\
 &c_{+}=c^2_V+c^2_A,\,\,\,c_{-}=c^2_V-c^2_A\,.
\label{E:2.4}
\end{align}
Note that the function $f_8$, given in Ref.~\cite{6}, contains an error
and we give here corrected expression for this function. It is
convenient in the further calculations to employ the relations,
which take place in the center-of-mass system:
\begin{align}
 &p^0_1=\frac{1}{2}q[1+\frac{m^2_a-m^2_b}{q^2}],\,\,\,
 p^0_2=\frac{1}{2}q[1+\frac{m^2_b-m^2_a}{q^2}],\notag\\
 &(p_1q)=\frac{1}{2}(q^2+m^2_a-m^2_b),\,\,\,
 (p_2q)=\frac{1}{2}(q^2+m^2_b-m^2_a),\notag\\
 &(p_1p_2)=\frac{1}{2}(q^2-m^2_a-m^2_b),\,\,\,|\vec{p}_1|=|\vec{p}_2|=
 \frac{1}{2}q\bar{\lambda}(m_a,m_b;q).
 \label{E:2.5}
 \end{align}
The analogous relations occur for the momenta $k_1$ and $k_2$
of the particles $c$ and $d$. In Eqs.(\ref{E:2.5}) the symbol $q$
has different meanings in the expressions $(p_1 q)$, $q=p_1+p_2$
(q is 4-momentum) and in the expression $q[1+f(q)]$, where
$q=\sqrt{(q\cdot q)}$ is a number.

With the help of the relations (\ref{E:2.2})-(\ref{E:2.5}) and above
discussed expressions for propagators,
we have got by tedious but straightforward calculations the
universal factorized cross-section for all permissible
combinations of particles $(a,b,R,c,d)$:
\begin{equation}
 \sigma(ab\rightarrow R\rightarrow cd)=\frac{16\pi (2J_R+1)}
 {(2J_a+1)(2J_b+1) \bar{\lambda}^2(m_a,m_b;\sqrt{s})}
 \frac{\Gamma^{ab}_R(s)\Gamma^{cd}_R(s)}{|P_R(s)|^2}.
\label{E:2.6}
\end{equation}
In Eq.(\ref{E:2.6})  $J_k$ is spin of the particle ($k=a,b,R$),
$s=(p_1+p_2)^2$, $\Gamma^{ab}_R(s)=\Gamma(R(s)\rightarrow ab)$ and
$P_R(s)$ is propagator's denominator of the UP or resonance $R$.
The expressions for $\Gamma^{ab}_R(s)$ and $\Gamma^{cd}_R(s)$
follow from Eqs.(\ref{E:2.2}-\ref{E:2.4}), when squared mass of UP
is $m^2_R=q^2=s$. The factorization of cross-section
does not depend on the definition of $P_R(s)$, which can be
determined in a phenomenological way, in Breit-Wigner or pole form
\cite{8,9}, etc. The expression (\ref{E:2.6}) is a natural
generalization of the spin-averaged Breit-Wigner (non-relativistic)
cross-section, defined by the expression (37.51) in Ref.
\cite{10}. Note that the factorization is exact in our approach,
while in the traditional one it occurs as an approximation.

The cross-section of exclusive process $ab\rightarrow
R\rightarrow cd$, defined by Eq.(\ref{E:2.6}), does not depend on
$J_c$ and $J_d$. So, it can be summarized over final channels
$R\rightarrow cd$:
\begin{equation}
 \sigma(ab\rightarrow R(s)\to all)=\frac{16\pi k_R}
 {k_a k_b\bar{\lambda}^2(m_a,m_b;\sqrt{s})}\frac{\Gamma^{ab}_R(s)\Gamma^{tot}_R(s)}
 {|P_R(s)|^2}.
 \label{E:2.7}
\end{equation}
In Eq.(\ref{E:2.7}) $k_i=2J_i+1$ and
$\Gamma^{tot}_R(s)=\sum_{cd}\Gamma^{cd}_R(s)$, where for
simplicity we restrict ourselves by two-particle channels.

The factorization effect, expressed by Eq.(\ref{E:2.6}), has two
aspects. On the one hand it means that the decay of UP proceeds
independently of its production in the approach considered. On the
other hand it leads to significant simplification of calculations,
in particular, in the case of the complicate processes
(the scattering with chain decay of products).

\section{Cross-section of the process $ab\rightarrow
R\rightarrow R_1 x\rightarrow cdx$}
Here, we consider factorization effects in the case of complicate chain processes. For example, let us discuss the process of scattering $ab\to R\to R_1 x$ with consequent decay $R_1\to cd$. In this case, Eq.~(\ref{E:2.6}) has the form:
\begin{equation}
 \sigma(ab\rightarrow R(s)\to R_1 x)=\frac{16\pi k_R}
 {k_a k_b\bar{\lambda}^2(m_a,m_b;\sqrt{s})}\frac{\Gamma^{ab}_R(s)\Gamma^{R_1x}_R(s)}
 {|P_R(s)|^2}.
 \label{E:3.1}
\end{equation}
 To calculate the value $\Gamma^{R_1 x}_R (s)$ we apply convolution formula, which accounts FWE in the decay $R(s)\to R_1 x$ \cite{6}:
\begin{equation}
 \Gamma(R(s)\to R_1 x)=\int_{q^2_1}^{q^2_2}\Gamma(R(s)\to R_1(q)x)\rho_{R_1}(q)\,dq^2\,.
 \label{E:3.2}
\end{equation}
In Eq.~(\ref{E:3.2}) $q=p_R-p_x$, $q_{1,2}$ are defined by kinematics of the process and $\rho_{R_1}(q)=q \Gamma^{tot}_{R_1}(q)/\pi|P_R(q)|^2$ is interpreted in the model \cite{6} as distribution function of the smeared mass of unstable particle $R_1$. Convolution structure of Eq.~(\ref{E:3.2}) is caused by factorization of decay rate $\Gamma(R\to R_1x\to x, all)$. This effect takes place exactly when the model propagators $\hat{\eta}(q)$ and $\eta_{\mu\nu}(q)$ are used (as in the present work).

From Eqs.~(\ref{E:3.1}) and (\ref{E:3.2}) it follows:
\begin{align}
 &\sigma(ab\rightarrow R\rightarrow R_1 x)=\notag\\&\frac{16\pi k_R}
 {k_a k_b\bar{\lambda}^2(m_a,m_b;\sqrt{s})}\frac{\Gamma^{ab}_R(s)}{|P_R(s)|^2}
 \int_{q^2_1}^{q^2_2}\Gamma(R(s)\rightarrow
 R_1(q)x)\rho_{R_1}(q)\,dq^2.
 \label{E:3.3}
\end{align}
Using the expression for $\rho_{R_1}(q)$, from Eq.~(\ref{E:3.3}) we can get   the cross-section of exclusive process, for example $ab\to R\to R_1 x\to cdx$. To this effect we represent $\Gamma ^{tot}_{R_1}(q)$ in the form:
\begin{equation}
 \Gamma^{tot}_{R_1}(q)=\sum_{X_1}\Gamma^{X_1}_{R_1}(q);\,\,\,
 \Gamma^{cd}_{R_1}(q)=\Gamma(R_1(q)\to cd)\,.
 \label{E:3.4}
\end{equation}
As a result, from (\ref{E:3.3}) and (\ref{E:3.4}) we get:
\begin{align}
 &\sigma(ab\rightarrow R\rightarrow cdx)=\notag\\
 &\frac{16k_R}{k_a k_b \bar{\lambda}^2(m_a,m_b;\sqrt{s})}\frac{\Gamma^{ab}_R(s)}{|P_R(s)|^2}
 \int_{q^2_1}^{q^2_2}\Gamma(R(s)\rightarrow
 R_1(q)x)\frac{q \Gamma^{cd}_{R_1}(q)}{|P_{R_1}(q)|^2}\,dq^2.
 \label{E:3.5}
\end{align}

Similar structure arises in the case $R\rightarrow R_1R_2$, i.e.
when there are two UP in the final state, which have
two-particle decay channels (semi-analytical approach). Thus, the
model gives a convenient instrument to describe two-particle
scattering accompanied by complicated decay-chain processes. However,
we have checked by direct calculations only two types of processes
- the decay of type $a\rightarrow Rx\rightarrow bcx$ \cite{6} and
the scattering of type $ab\rightarrow R\rightarrow cd$. The more
complicated processes, such as decay $a\rightarrow
R_1 R_2\rightarrow cdef$ and scattering $ab\rightarrow
R\rightarrow R_1 R_2\rightarrow cdef$, will be the subject of
the next paper.

\section{Methodological and phenomenological\\ analysis of the factorization effect}

The model factorization of a decay width and cross-section of the
processes with UP in an intermediate state was established by
straightforward calculations at tree level. However, these
calculations in the effective theory of UP \cite{7} account for
some loop diagrams. The vertex and self-energy type corrections can be included into $\Gamma_R(s)$ and $P_R(s)$ respectively. These corrections do
not breakdown a factorization, but the interaction between initial
and final states does. However, such an interaction has no clear
and explicit status in perturbation theory due to UP (or
resonance) is not a perturbative object in the resonance
neighborhood \cite{7,11,11a}. As it was noted in Ref.~\cite{12}, such
non-factorizable corrections give small contribution to the processes
$e^{+}e^{-}\to ZZ, WW, 4f$ near the resonance range.

Now we consider another aspect of factorization effect, namely, the
determination of dressed propagator of UP. Factorization of decay
width and cross-section does not depend on the structure of
propagator's denominator $P_R(q)$, but crucially depends on the
structure of its numerator in the case of vector and spinor UP. As
it was verified by direct calculations, the factorization always
takes place in the case of scalar UP. The expressions
$\eta_{\mu\nu}(m_R) =g_{\mu\nu}-q_{\mu}q_{\nu}/m^2_R$ and
$\hat{\eta}(m_R)=\hat{q}+m_R$ for vector and spinor UP,
respectively, do not lead to exact factorization. But the expressions
$\eta_{\mu\nu}(q) =g_{\mu\nu}-q_{\mu}q_{\nu}/q^2$ and
$\hat{\eta}(q)=\hat{q}+q$ strictly lead to factorization for any
kinds of other particles. It should be noted that the definition
of the functions $\eta_{\mu\nu}(q)$ and $\hat{\eta}(q)$ is not related with
the choice of the gauge, because effective theory of UP \cite{7} is
not the gauge theory. The choice of $q$ instead of $m_R$ in the
$\eta_{\mu\nu}$ and $\hat{\eta}$ may seems contradict to the
equation of motion for vector and spinor UP. However, this
statement is valid for the stable particle with fixed mass. In the
case of UP the question arises what the mass participates in
equation of motion - pole mass or one of the renormalized
mass \cite{13}? An account of uncertainty relation by smearing of
mass intensifies the question. There is no unique and strict
determination of dressed propagator structure for vector and
spinor UP due to the specific nature of Dyson summation in these cases
\cite{6}. The situation is more complicated and involved in the
case of hadron resonance. So, the functions $\eta_{\mu\nu}$ and
$\hat{\eta}$ have rather phenomenological (or model) than
theoretical status. The model of UP \cite{7} defines these
functions as $\eta_{\mu\nu}(q)$ and $\hat{\eta}(q)$, which
describe the dressed propagators of UP in the resonance
neighborhood.

Further, we briefly analyze the phenomenological aspect of
factorization. In the low-energy experiments of type
$e^+e^-\rightarrow \rho, \omega...\rightarrow \pi^+\pi^-, ...$ we
can not distinguish propagators $\eta_{\mu\nu}(m_R)$ and
$\eta_{\mu\nu}(q)$ even for the wide resonance. This is due to
the equality $\bar{e}^-(p_1)(\hat{p}_1+\hat{p}_2)e^-(p_2)=0$, when the
functions $\eta_{\mu\nu}$ reduce to $g_{\mu\nu}$ in both cases.
In the high-energy experiments of type $e^+e^-\rightarrow
Z\rightarrow \bar{f}f$, where $f$ is quark or lepton (we neglect
$\gamma-Z$ interference), the transverse part of amplitude is
\begin{equation}
 M_q\sim\bar{e}^-(p_1)\hat{q}(c_e-\gamma_5)e^-(p_2)\bar{f}^+(k_1)\hat{q}
 (c_f-\gamma_5)f^+(k_2),
\label{E:4.1}
\end{equation}
where $q=p_1+p_2=k_1+k_2$. From Eq.(\ref{E:4.1}) with the help of
the Dirac equations in momentum representation it follows
\begin{equation}
 M_q\sim m_em_f \bar{e}^-(p_1)\gamma_5
 e^-(p_2)\bar{f}^+(k_1)\gamma_5f^+(k_2).
 \label{E:4.2}
\end{equation}
As a result, we get the terms $m_e m_f/q^2$ and $m_e m_f/m^2_Z$ for $\eta(q)$ and $\eta(m_R)$,
respectively. The  difference of these values is of the order of $(m_e m_f/m^2_Z)\cdot(m_Z-q)/m_Z$
at energy $q^2\sim m^2_Z$. Thus, the distinction between the structure of two type
of the expressions $\eta_{\mu\nu}$ is negligible in a wide range of energy.

The structure of $\hat{\eta}$ can be studied in the process of type
$VF\rightarrow R\rightarrow V^{'}F^{'}$, where $V$ and $F$ are vector and fermion field, $R$ is, for instance, baryon resonance with a large
width. In this case, the difference between $\hat{\eta}(m_R)$ and
$\hat{\eta}(q)$ is characterized by the value $\sim\Gamma_R/m_R$
at peak region, and this problem demands more detailed
analysis.

From this analysis it follows that method of factorization is a simple analytical analog of narrow-width approximation (NWA, which contains five critical assumptions \cite{5}). Instead, we use the structure of propagators' numerators $\eta(q)$, which follows from usual ones under a simple transformation $m_R \to q$, and one assumption: there is no significant interference with non-resonant processes (fifth assumption of NWA). The rest assumptions of NWA can be derived from the first our point, where some of them are not obligatory in the special cases. The method leads to factorization in two type of processes - in the decay-chain \cite{6} (universal convolution formula) and scattering ones (universal formula (\ref{E:2.6})). Combining these two results, we get a simple and strict algorithm of analytical description of the complicated processes.

\section{Conclusion}

The factorization effect gives us a convenient phenomenological way
to describe the three-particle decays and two-particle scattering
processes. This effect significantly simplifies calculations and
gives compact universal formulae for the decay rate and
cross-section.

In this work, we have shown that the factorization always is
valid when scalar UP is in the intermediate state. In the case of
vector or spinor intermediate states, the factorization takes
place when the specific propagators are used for these states. These
propagators are derived in the model of UP with a random (smeared) mass.
They negligibly differ from the traditional propagators at peak area and
follow from the smearing of mass in accordance
with the uncertainty relation. Our method makes it
possible significantly simplify the calculation of the complicated
decay-chain and scattering processes. It is some analytical analog of NWA
and gives a simple and strict algorithm for calculations. This approach can be
treated also as convenient approximation, which always is valid in the
resonance range, where non-resonance contribution is small.

We have fulfilled also a short methodological and phenomenological
analysis of the approach under discussion. It was shown, that in
the process $e^+e^-\rightarrow f\bar{f}$ the difference between
two forms of propagators is negligible in a wide range of energy. It can be significant in the processes with baryon resonance in an intermediate state, but in this case we should fulfill an additional analysis.

\section{Appendix}

In this section, we briefly describe the model of UP with a smeared mass and construct the propagators for the vector and spinor fields. The structure of these propagators lead to the factorization effect in the processes with the participation of the UP in the intermediate state. The model field wave function, which describes UP, is represented in the form \cite{7}:
\begin{equation}\label{A.1}
 \Phi_a(x)=\int\Phi_a(x,\mu)\omega(\mu)d\mu,
\end{equation}
where $\Phi_a(x,\mu)$ is standard spectral component, which defines a particle
with a fixed mass squared $m^2=\mu$ in the stable particle
approximation (SPA). The weight function $\omega(\mu)$ is
formed by the self-energy interactions of UP with vacuum fluctuations
and decay products. This function describes the smeared (fuzzed)
mass-shell of UP.

The model Lagrangian, which determines a "free" (effective) unstable field
$\Phi(x)$, has the convolution form:
\begin{equation}\label{A.2}
 L(\Phi(x))=\int L(\Phi(x,\mu))|\omega(\mu)|^2\,d\mu\,.
\end{equation}
In Eq.(\ref{A.2}) $L(\Phi(x,\mu))$ is the standard Lagrangian,
which describes model "free" field component $\Phi(x,\mu)$ in the stable
particle approximation ($m^2=\mu$).

From Eq.(\ref{A.2}) and prescription
$\partial\Phi(x,\mu)/\partial\Phi(x,\mu^{'})=\delta(\mu-\mu^{'})$
it follows the Klein-Gordon equation for the spectral component of
scalar or vector field:
\begin{equation}\label{A.3}
 (\square-\mu)\Phi_{\alpha}(x,\mu)=0.
\end{equation}
In analogy with (\ref{A.3}) one can get the Dirac equation for
fermion spectral component. As a result, we get the  standard
representation of the field function $\Phi_{\alpha}(x,\mu)$ with a fixed mass parameter
$\mu$ (spectral component).
 All standard definitions, relations and frequency expansion take
place for $\Phi_{\alpha}(k,\mu)$, however, the relation
$k^0_{\mu}=\sqrt{\bar{k}^2+\mu}$ defines the smeared (fuzzy)
mass-shell due to a random nature of the mass parameter $\mu$. The convolution (diagonal)
representation of the "free" Lagrangian (\ref{A.2}) has an
assumption (or approximation?) that the states with different $\mu$
do not interact in the approximation of the model "free" fields.

The expressions (\ref{A.1})--(\ref{A.3}) define the model
"free" unstable field as some effective field. As it was mentioned
above, this field is formed by an interaction of "bare" UP with
the vacuum fluctuations and decay products, that is includes
self-energy contribution in the resonant region. Such an
interaction leads to the spreading (smearing) of mass, described by
the function $\omega(\mu)$ or $\rho(\mu)=|\omega(\mu)|^2$. Thus, we go from the
distribution $\rho^{st}(\mu)=\delta(\mu-M^2)$ for "bare" particles
to some smooth density function $\rho(\mu)=|\omega(\mu)|^2$ with
mean value $\bar{\mu}\approx M^2$ and mean square deviation
$\sigma_{\mu}\approx \Gamma$. So, the UP is characterized by the
weight function $\omega(\mu)$ or probability density $\rho(\mu)$
with parameters $M$ and $\Gamma$ (or real and imaginary parts of a
pole).

The commutative relations for the model operators have an additional
$\delta$-function:
\begin{equation}\label{A.4}
 [\dot{\Phi}^{-}_{\alpha}(\bar{k},\mu),\,\Phi^{+}_{\beta}(\bar{q},\mu^{'})]_{\pm}
 =\delta(\mu-\mu^{'}) \delta(\bar{k}-\bar{q})\delta_{\alpha\beta},
\end{equation}
where subscripts $\pm$ correspond to the fermion and boson fields.
The presence of $\delta(\mu-\mu^{'})$ in Eq.(\ref{A.4}) means an
assumption - the acts of creation and annihilation of the
particles with various $\mu$ (the random mass squared) do not
interfere. Thus, the parameter $\mu$ has the status of physically
distinguishable value of a random $m^2$. This assumption is
naturally related with a diagonal form of Eqs.(\ref{A.2}) and
(\ref{A.3}) and directly follows from the interpretation of $q^2$
as a random parameter $\mu$. By integrating the
both sides of Eq.(\ref{A.4}) with weights
$\omega^{*}(\mu)\omega(\mu^{'})$ one can get the standard
commutative relations
\begin{equation}\label{A.5}
 [\dot{\Phi}^-_{\alpha}(\bar{k}),\Phi^+_\beta(\bar{q})]_{\pm}=\delta(\bar{k}-\bar{q})
 \delta_{\alpha\beta}\,,
\end{equation}
where $\Phi^{\pm}_{\alpha}(\bar{k})$ is the full operator field
function in the momentum representation
\begin{equation}\label{A.6}
 \Phi^{\pm}_{\alpha}(\bar{k})=\int\Phi^{\pm}_{\alpha}(\bar{k},\mu)\omega(\mu)d\mu\,.
\end{equation}
It should be noted that Eq.(\ref{A.6}) follows from
Eq.(\ref{A.5}) when $\int|\omega(\mu)|^2d\mu=1$, that is
$|\omega(\mu)|^2$ can be interpreted as a normalized probability density.

The expressions (\ref{A.1}), (\ref{A.2}) and (\ref{A.4}) are the
principal elements of the model. The weight function
$\omega(\mu)$ (or $\rho(\mu)$) is full
characteristic of UP in the framework of the model. The relations (\ref{A.4}) define the
structure of the model amplitude and transition probability.

Here, we consider the model amplitude for the simplest
processes with UP in an initial or final state and get the convolution formula as a
direct consequence of the model. The expression for a scalar
operator field \cite{7} is
\begin{equation}\label{A.7}
 \phi^{\pm}(x)=\frac{1}{(2\pi)^{3/2}}\int\omega(\mu)d\mu\int\frac{a^{\pm}(\bar{q},\mu)}
 {\sqrt{2q^0_{\mu}}}e^{\pm iqx}d\bar{q}\,,
\end{equation}
where $q^0_{\mu}=\sqrt{\bar{q}^2+\mu}$ and $a^{\pm}(\bar{q},\mu)$
are the creation or annihilation operators of UP with the momentum $q$ and
mass squared $m^2=\mu$. Taking into account Eq.(\ref{A.4}) one
can get:
\begin{equation}\label{A.8}
 [\dot{a}^{-}(\bar{k},\mu),\phi^{+}(x)]_{-};\,\,\, [\phi^{-}(x), \dot{a}^{+}(\bar{k},\mu)]_{-}
 =\frac{\omega(\mu)}{(2\pi)^{3/2}\sqrt{2k^0_{\mu}}}e^{\pm ikx}\,,
\end{equation}
where $k^0_{\mu}=\sqrt{\bar{k}^2+\mu}$. The expressions (\ref{A.8}) differ from the standard ones by the
factor $\omega(\mu)$ only. From this result it follows that, if
$\dot{a}^{+}(k,\mu)|0\rangle$ and $\langle0|\dot{a}^{-}(k,\mu)$
define UP with the mass $m=\sqrt{\mu}$ and momentum $k$ in the
initial or final states, then the amplitude for the transition
$\Phi\rightarrow\phi\phi_1$ is
\begin{equation}\label{A.9}
 A(k,\mu)=\omega(\mu)A^{st}(k,\mu)\,,
\end{equation}
where $A^{st}(k,\mu)$ is the amplitude in the stable particle
approximation. This amplitude is calculated in the standard way
and can include the higher corrections. Moreover, it can be an
effective amplitude for the processes with hadron participation.
From Eq.(\ref{A.9}) it follows that the
 differential (on $\mu$) probability of transition is
$dP(k,\mu)=\rho(\mu)|A(k,\mu)|^2d\mu$.

To define the transition probability of the process
$\Phi\rightarrow\phi\phi_1$, where $\phi$ is UP with a large
width, we should take into account the status of the parameter $\mu$
as a physically distinguishable value, which follows from
Eq.(\ref{A.4}). Thus, the differential (on $k$) probability is
\begin{equation}\label{A.10}
 d\Gamma(k)=\int d\Gamma^{st}(k,\mu)\rho(\mu)d\mu\,.
\end{equation}
In Eq.(\ref{A.10}) the differential probability
$d\Gamma^{st}(k,\mu)$ is defined in the standard way (the stable
particle approximation):
\begin{equation}\label{A.11}
 d\Gamma^{st}(k,\mu)=\frac{1}{2\pi}\delta(k_{\Phi}-k_{\phi}-k_1)|A^{st}
 (k,\mu)|^2d\bar{k}_{\phi}d\bar{k}_1\,,
\end{equation}
where $k=(k_{\Phi},k_{\phi},k_1)$ denotes the momenta of
particles. From Eqs.(\ref{A.10}) and (\ref{A.11}) it directly
follows the well-known convolution formula for a decay rate
\begin{equation}\label{A.12}
 \Gamma(m_{\Phi},m_1)=\int_{\mu_1}^{\mu_2}\Gamma^{st}(m_{\Phi},m_1;\mu)\rho(\mu)d\mu\,,
\end{equation}
where $\rho(\mu)=|\omega(\mu)|^2$, $\mu_1$ and $\mu_2$ are the
threshold and maximal invariant mass squared of an unstable
particle $\phi$.

An account of higher corrections in the amplitude (\ref{A.9})
keeps the convolution form of Eq.(\ref{A.12}). This form can be
destroyed by the interaction between the products of UP ($\phi$)
decay and initial $\Phi$ or final $\phi_1$ states. The calculation
in this case can be performed in the standard way, but UP in the
intermediate state is described by the model propagator. However,
the calculation within the framework of perturbative theory (PT)
can not be applicable to the UP with a large width, that is to the
short-living particle. In any case, the applicability of the PT,
of the model approach or convolution method to the decays considered
should be justified by an experiment. The validity of the CM was demonstrated
for many processes, but this problem needs in more detailed
investigation.
If there are two UP with large widths in a final state
$\Phi\rightarrow\phi_1\phi_2$, then in analogy with the previous
case one can get the double convolution formula:
\begin{equation}\label{A.13}
 \Gamma(m_{\Phi})=\int\int\Gamma^{st}(m_{\Phi};\mu_1,\mu_2)\rho_1(\mu_1)\rho_2(\mu_2)d\mu_1
 d\mu_2\,.
\end{equation}
The derivation of CF for the cases when there is a vector or
spinor UP in the final state can be done in analogy with the case
of scalar UP. However, in Eqs.(\ref{A.7}), (\ref{A.8}) and
(\ref{A.9}) one should take into account the polarization vector
$e_m(q)$ or spinor $u^{\nu,\pm}_{\alpha}(q)$, where momentum $q$ is on fuzzy
mass-shell. As a result, we get the polarization matrix with
$m^2=\mu$. In the case of vector UP in the final state we have
\begin{equation}\label{A.14}
 \sum_{e} e_m(q)e^{*}_n(q)=-g_{mn}+q_mq_n/\mu\,,
\end{equation}
and in the case of spinor UP in the final state:
\begin{equation}\label{A.15}
 \sum_{\nu} u^{\nu,\pm}_{\alpha}(q)\bar{u}^{\nu,\mp}_{\beta}(q)=\frac{1}{2q^0_{\mu}}
 (\hat{q}\mp\sqrt{\mu})_{\alpha\beta}\,,
\end{equation}
where the summation over polarization is implied and
$q^0_{\mu}=\sqrt{\bar{q}^2+\mu}$. The same relations take place for the initial states, however one have to
average over the polarizations. The formulae (\ref{A.12}) and (\ref{A.13}) describe FWE in full
analogy with the phenomenological convolution method. Similar
method, called the semi-analytical approach, was applied in
calculations of cross-section of the processes $e^+e^-\rightarrow
ZZ,WW$ at LEP2 energy \cite{13}, where the phase space of the final
states was integrated in analogy with (\ref{A.13}). This approach
gives a simple expressions for the cross-sections, which are
equivalent to the inclusive cross-section of the fourth-fermion reactions.
The calculation of this cross-section in the framework of
standard PT is very complicated and usually carried out with the
help of the Monte-Carlo simulation. The model under consideration gives a
quantum field basis for CM, which takes into account the
fundamental uncertainty relation, provides a simple expressions for decay rates and
is in a good agreement with the experimental data on some processes. To evaluate FWE in the
case, when UP is in an initial state, we have to take into
consideration the process of UP production. If UP is in an
intermediate state, then the description of FWE is equivalent to
the traditional one, but the propagators are determined by the model.

Now, we consider the the structure of the model propagators.
With the help of the traditional method, one can get from
Eqs.(\ref{A.1}), (\ref{A.4}) and (\ref{A.6}) the expression for
the unstable scalar Green function \cite{7}:
\begin{equation}\label{A.16}
 \langle 0|T(\phi(x),\phi(y))|0\rangle\equiv D(x-y)=\int
 D(x-y,\mu)\rho(\mu)d\mu\,.
\end{equation}
In Eq.(\ref{A.16}) $D(x,\mu)$ is a standard scalar Green function
with $m^2=\mu$, which describes UP in an intermediate state:
\begin{equation}\label{A.17}
 D(x,\mu)=\frac{i}{(2\pi)^4}\int\frac{e^{-ikx}}{k^2-\mu+i\epsilon}dk\,.
\end{equation}
The right-hand side of Eq.(\ref{A.16}) is the Lehmann-like
spectral (on $\mu$) representation of the scalar Green function. Taking into
account the relation between scalar and vector Green functions, we
can get the Green function of the vector unstable field in the
form:
\begin{align}\label{A.18}
 D_{mn}(x,\mu)=&-(g_{mn}+\frac{1}{\mu}\frac{\partial^2}{\partial
 x^n\partial x^m})D(x,\mu)\notag\\
 =&\frac{-i}{(2\pi)^4}\int\frac{g_{mn}-k_m
 k_n/\mu}{k^2-\mu+i\epsilon}e^{-ikx}dk\,.
\end{align}
Analogously, the Green function of the spinor unstable field is
\begin{equation}\label{A.19}
 \Hat{D}(x,\mu)=(i\hat{\partial}+\sqrt{\mu})D(x,\mu)=\frac{i}{(2\pi)^4}
 \int\frac{\hat{k}+\sqrt{\mu}}{k^2-\mu+i\epsilon}e^{-ikx}dk\,,
\end{equation}
where $\hat{k}=k_i\gamma^i$. These Green functions in momentum
representation have a convolution form:
\begin{equation}\label{A.20}
 D_{mn}(k)=\int D_{mn}(k,\mu)\rho(\mu)d\mu\,,\,\,\,\,\,
 \Hat{D}(k)=\int \Hat{D}(k,\mu)\rho(\mu)d\mu\,.
\end{equation}
The expression (\ref{A.18}-\ref{A.20}) for the propagators of vector and spinor fields
leads to the effect of factorization, which in turn, gives the convolution formula.

\end{document}